\documentclass[aps,pre,twocolumn,groupedaddress]{revtex4-1}

\usepackage{graphicx}
\usepackage{amsmath}
\usepackage{amssymb}
\usepackage{amsthm}
\usepackage{arydshln} 
\usepackage{tikz}
\usetikzlibrary{shapes}
\usetikzlibrary{arrows}
\usepackage{color}

\begin{document}

\newtheorem{corollary}{Corollary}[section]
\newtheorem{lemma}{Lemma}[section]
\newtheorem{theorem}{Theorem}[section]
\newtheorem{definition}{Definition}[section]


\title{Citation Analysis with Mark-and-Recapture}


\author{Chuan Wen, Loe$^1$ and Henrik Jeldtoft Jensen$^2$}
\affiliation{}


\date{\today}

\begin{abstract}
Mark-and-Recapture is a methodology from Population Biology to estimate the number of a species without counting every individual. This is done by multiple samplings of the species using traps and discounting the instances that were caught repeated. In this paper we show that this methodology is applicable for citation analysis as it is also not feasible to count all the relevant publications of a research topic. In addition this estimation also allows us to propose a stopping rule for researchers to decide how far one should extend their search for relevant literature.
\end{abstract}

\pacs{}

\maketitle

\section{Introduction}
There are many situations where one cannot explicitly count all the instances to determine the size of a population, e.g. the number of polar bears in Western Canadian Arctic \cite{demaster1980multiple}. Hence to estimate the population size, a statistical sampling method known as \emph{Mark-and-Recapture} is used in Population Biology \cite{southwood2009ecological}.

This statistical approximation is not limited to ecology and can be applied to epidemiology \cite{chao2001applications}, linguistics \cite{Alcoy} and software engineering \cite{chao1993stopping}. In essence Mark-and-Recapture measures the completeness of a sampling over a set. Hence we applied this methodology to assess the completeness of the bibliography of literature reviews.

A literature review is a summary of a research topic where its source of information is curated by domain experts. The authors often have to rely on specialized search engines like Google Scholar, Microsoft Academic Search, or Web of Science to find all the relevant publications. However the number of results from these search engines can easily be in the order of hundreds of thousands, and most researchers rely on their gut feelings to stop their search.

This is a similar problem faced by clinical researchers as the results of medical trials are disparate in different databases (Medline, EMBASE, CINAHL, and EBM reviews). Thus clinical researchers used Mark-and-Recapture as a stopping rule to estimate the completeness of their research \cite{lane2013capture,kastner2009capture}. In this paper we extend the idea to a different disciplines and assess the quality of different academic search engines.


\section{Population Estimation}
It is highly probable that the bibliography of the literature reviews are incomplete. Just like population biology, it is not possible to capture all the animals to determine the population of an animal species. Hence \emph{mark-and-recapture} can be used to approximate the population by sampling the species repeated and discount for the number of instances that were caught  previously. 

\subsection{Mark-And-Recapture}
Animals are captured and marked before releasing them back in the wild. After enough time has pass to allow a complete mixing, the population is sampled for the second time. In the second sample, the ratio of marked animals (from the first capture) to the number of captured animals is approximately the ratio of captured animals in the first sample to the total population, hence by 
the Peterson method \cite{southwood2009ecological}:
\begin{equation}\label{eq:peterson}
\mbox{Total population} \approx \frac{N_1N_2}{R},
\end{equation}
with standard deviation
\begin{equation}\label{eq:peterson_stddev}
\sigma=\sqrt{\frac{(N_1+1)(N_2+1)(N_1-R)(N_2-R)}{(R+1)^2(R+2)}}, 
\end{equation}
where $N_1$, $N_2$ is the number of captures in the $1^{st}$ and $2^{nd}$ sample respectively, and $R$ is the number of marked animals (individuals that were captured in both samplings). Furthermore Mark-And-Recapture can be extended to multiple captures by a weighted average of Eq. \ref{eq:peterson} known as Schnabel Index \cite{schnabel1938estimation}:
\begin{equation}\label{eq:multiple_peterson}
\mbox{Total population} \approx \frac{\sum_{i=1}^{m}N_iM_i}{\sum_{i=1}^{m}R_i},
\end{equation}
with standard deviation
\begin{equation}\label{eq:multiple_peterson_stddev}
\sigma=\sqrt{\frac{\sum_{i=1}^{m}R_i}{(\sum_{i=1}^{m}N_iM_i)^2}}, 
\end{equation}
where $N_i$ is the number of captures in the $i^{th}$ sample, $M_i$ is the 
total number of marked animals in the population before the $i^{th}$ sample, 
and $R_i$ is the number of marked captures in the $i^{th}$ sample.

\subsection{Assumptions in the Estimation}\label{section:lr_assumptions}
To apply the same methods to scientific literature analysis, the assumptions have to be parallel to Population Biology. The mixing period for population biology has to be long enough such that the second sampling is independent from the first, yet short enough to minimize the effects of population changes or the death of the tagged animals, i.e. a closed population. Hence the literature reviews have to be independent efforts and completed around the same time. This assumption is easily true for the results from different independent search engines.

However the probability that a paper is found and referenced is not equal \cite{bennett2004capture}. There are many factors that affects the visibility of a publication in a search engine (respectively literature review), e.g. quality of research, discipline, keywords, date of publication, authors, etc. This is a common violation of assumption in wildlife as some animals have a higher tendency to be captured again, i.e. ``trap-happy" animals. In such cases, we can assume the result is the lower bound to the true population size.

\section{Comparing the Bibliography of Literature Reviews}
\subsection{Experiment Methodology}
There are several reviews on the communities detection algorithms of graphs over the past decade --- Newman 2004 \cite{newman2004detecting}, Fortunato and Castellano 2007 \cite{fortunato07}, Schaeffer 2007 \cite{schaeffer2007graph}, Porter \emph{et al.} 2009
 \cite{porter2009communities}, and Fortunato 2010 \cite{fortunato2010community}. Although it is tempting to apply Schnabel Index to sample of the body of literature repeatedly, it violates many assumptions of the estimator which will make the results questionable.

The first violation is that these surveys are not independent sampling of the literature as most of them cited the earlier reviews. Secondly the population in question is not closed as there are many publication on communities detection since year 2004. There is only 44 references in Newman 2004 review versus the 457 references in the review by Fortunato in 2010. Thus the results will be meaningless even if the numbers appears to support the methodology. 

Therefore to minimize the violation of the assumptions, the reviews must be published approximately the same year and the latter should not cite the earlier review. Hence in this case Schaeffer 2007 will the first sample and the review by Fortunato and Castellano 2007 will be the second. Finally the result will be compared against the bibliography of the review by 
Fortunato 2010 to gauge the accuracy of this methodology.

\subsection{Results}
Out of the 249 references in Schaeffer 2007, only 43 articles are directly relevant to communities detection. Most of the excluded references are on graph cutting from graph theory or clustering algorithms from machine learning as they do not connote the idea of modularity of communities in the articles. Similarly only 55 articles are chosen from the 97 references in the review by Fortunato and Castellano 2007.

Finally since there are only 20 relevant citations that were listed in both reviews, Eq. \ref{eq:peterson} and Eq. \ref{eq:peterson_stddev} suggest that there are $\approx 118 \pm 14$ publications on graph communities by 2007. In comparison, there are 112 articles before 2008 on graph communities in the bibliography of Fortunato 2010. The agreement is surprisingly good and supports the framework to use Mark-And-Recapture to determine the completeness of a literature review.

\section{Comparing Search Engines}\label{sec:compareSE}
Since literature reviews are well curated, the estimate from Mark-And-Recapture may suggests the size of the body of literature on a given topic. It gives new researchers a level of confidence in their preliminary investigations.

However the conditions for this methodology are hard to meet (section \ref{section:lr_assumptions}) for most research topics. Furthermore it begs the question, \emph{are the bibliography of the literature reviews complete?}. Since academic search engines are the basic sources of information for researchers, it is interesting to apply Mark-And-Recapture to compare the results from the different search engines.

\subsection{Related Work}
The preliminary process of a research is the task of searching and \emph{re-searching} the relevant publications to provide a comprehensive overview of a topic. There is no optimal stopping rule to determine if one has done sufficient search for the relevant articles, especially prolonged search will eventually reach a point of diminishing returns. This is a foremost challenge for any researchers and one of the reasons for peer reviewing publications (i.e. to avoid duplicated research).

The right balance between the time needed to find the relevant materials versus the quantity of materials is of particular interests for medical research. Given the growing amount of research versus the urgency to provide the proper medical care, the time spent on research have to be optimized. However the citation network of related clinical trials is disconnected, which reflects the possibility that the ``different camps" of clinical researchers use different research tools and hence unaware of the relevant literature from the other ``camps" \cite{robinson2014citation}. 

Thus Mark-And-Recapture methodology was proposed as a stopping rule for medical medical research \cite{lane2013capture,stelfox2013capture,kastner2009capture,kastner2007estimating,
booth2010much}. For example the empirical evaluation on osteoporosis disease management publication estimates approximately 592 articles are missing from 4 main bibliographic databases --- MEDLINE, EMBASE, CINAHL, and EBM Reviews \cite{kastner2007estimating}.

\subsection{Experiment Methodology}
The above framework however cannot be easily adopted for the researchers in other fields of science. Many keywords in science have multiple meanings in different contexts, for example the word \emph{graph} can be defined as a plot of a function or an abstract mathematical object. Hence there can be many unrelated results and thus the search engine can easily return hundreds of thousands of articles.

One way to sieve through the articles is to accept the ``top" relevant articles (suggested by the search engine) until no new significant information is gained \cite{browne2007cognitive}. However the measure of \emph{information gain} cannot be quantified and is often based on subjective gut feelings. We address this issue with an application of Mark-And-Recapture on the following academic search engines: Google Scholar, Microsoft Academic Search, or Web of Science.

The web-crawler and the database of these search engines are the ``traps" for the entire body of literature, and the ordering of the results is a reflection of the (search engine) algorithms' unique perspectives of the keywords. Suppose the top $n^{th}$ results of two search engines, $E_1$ and $E_2$, have $R$ number of common articles. Eq. \ref{eq:peterson} suggests that there are at least a total of $T=n^2/R$ publications on this topic. To avoid division by zero, we initialized $R=1$ at the beginning.

If we assumed that when one stops at the $n^{th}$ entry of $E_1$ and $E_2$, then the coverage of the body of literature is at most $C=(2n-R)/T$. Therefore the rate of change of $C$ with respect to $n$ estimates the information gained during the time spent with the search engines. A low rate of change implies low information gain and quantifies a stop to the search.

For simplicity, this paper only compares two search engines at a time where each of them are independent samplings over the body of literature. The ordering of the results is sorted by ``relevance" which is ranked by the different algorithms of the search engines. 

Lastly only the top 500 results from each search engine are collected in the experiments since Web of Science limits that number of articles to be exported at each time. Moreover if the sampling is too large it will trigger Google Scholar to temporarily ban users from accessing to its database. The software used to extract from Google Scholar and Microsoft Academic Search is \emph{Publish or Perish} \cite{pop4}.

\subsection{The Results from the Comparisons of the Search Engines}
In the experiments, we noticed that some papers are published in multiple sources, e.g. arXiv and peer-review journals. This will cause the search engines to occasionally return the same paper as multiple and distinct publications. Since there is no information gain for repeated articles, we have to adjust our equations.

The coverage $C$ of a literature can be viewed as a time series where the $n^{th}$ unit of time refers to the $n^{th}$ article of the search engines. Let $N_{i,n}$ be the number of \emph{unique} articles returned by search engine $E_i$ at time $n$. If $T$ is the estimated total number of publications on this topic at time $n$, then Eq. \ref{eq:peterson} gives us $T=N_{1,n}N_{2,n}/R$ where $R$ is the number of unique articles that are found in both search engines. Similarly, the coverage of the body of literature has to be adjusted as $C=(N_{1,n}+N_{2,n}-R)/T$.

In most cases $N_{1,n}=N_{2,n}\approx n$, which is the easiest to analyze. If $R$ converges to a constant, then $\lim_{n\rightarrow\infty}C \approx 1/n \rightarrow 0$. This implies that the further you continue searching with the same keywords, there is a diminishing returns to the information gain. This is more obvious if we consider the plot of $T$ as a function of $n$.

If $R$ converges to a constant, then $\lim_{n\rightarrow\infty}T \approx n^2 \rightarrow \infty$. This implies that the given keyword is so imprecise that the results from different search engines diverge as there is almost no common articles between the search engines.

In contrast if the rate of growth of $R$ is linearly bounded by $n$, then there is at most $n$ common articles at time $n$. Although the coverage $C\approx1.0$ and the estimated total number of articles is $n$, the figures are not meaningful. This is because it implies that the results of $E_1$ and $E_2$ are so similar that it is analogous to using only one search engine. In such case we are back to the original situation where we do not have a quantified method to analyze research results. 

Fortunately $R$ generally do not grow linearly for the entire time series and can be analyzed by plotting $T$ as a function of $n$. In fact $R$ tends to be sublinear and the coverage will approach zero. Hence the optimal stopping rule is to stop at a point when the derivative of $C$ is zero, this implies that the search has diminishing returns.

At the local maximum of $C$, further search have negative returns as the search engines' perspectives of the keyword begin to diverge. This is supported by the quadratic growth of $T$ after the stopping point. Hence the reason to stop is that the subsequent articles is less relevant from the perspective of the other search engines.

At the local minimum of $C$, the stopping rule is slightly counter-intuitive. As the coverage increases, technically it is prudent to continue the search as it implies that the researcher have more complete coverage of the literature. However for the coverage to increase rapidly, $R$ has to increase rapidly too. It usually means that the subsequent articles are already returned in the earlier results, and hence no information gain.

Finally if $N_{1,n}\neq n$, then $T$ is sublinear. This implies that one of the search engine listed the same publications from different journals/sources. By definition, two articles are the same if they have the same title and authored by the same people.

\subsection{Empirical Results}
The keywords chosen for the experiments are primarily based on our familiarity with the topics in Physics and Computer Science. The remaining keywords from the other disciplines are selectively chosen from ScienceWatch.com publication on the top 100 key scientific research front for 2013  \cite{king2013research}. The results are categorized such that each type has different stopping rules. 

\subsubsection{Type I (Convergence to Zero)}
\begin{figure}
  \includegraphics[width=1.0\columnwidth]{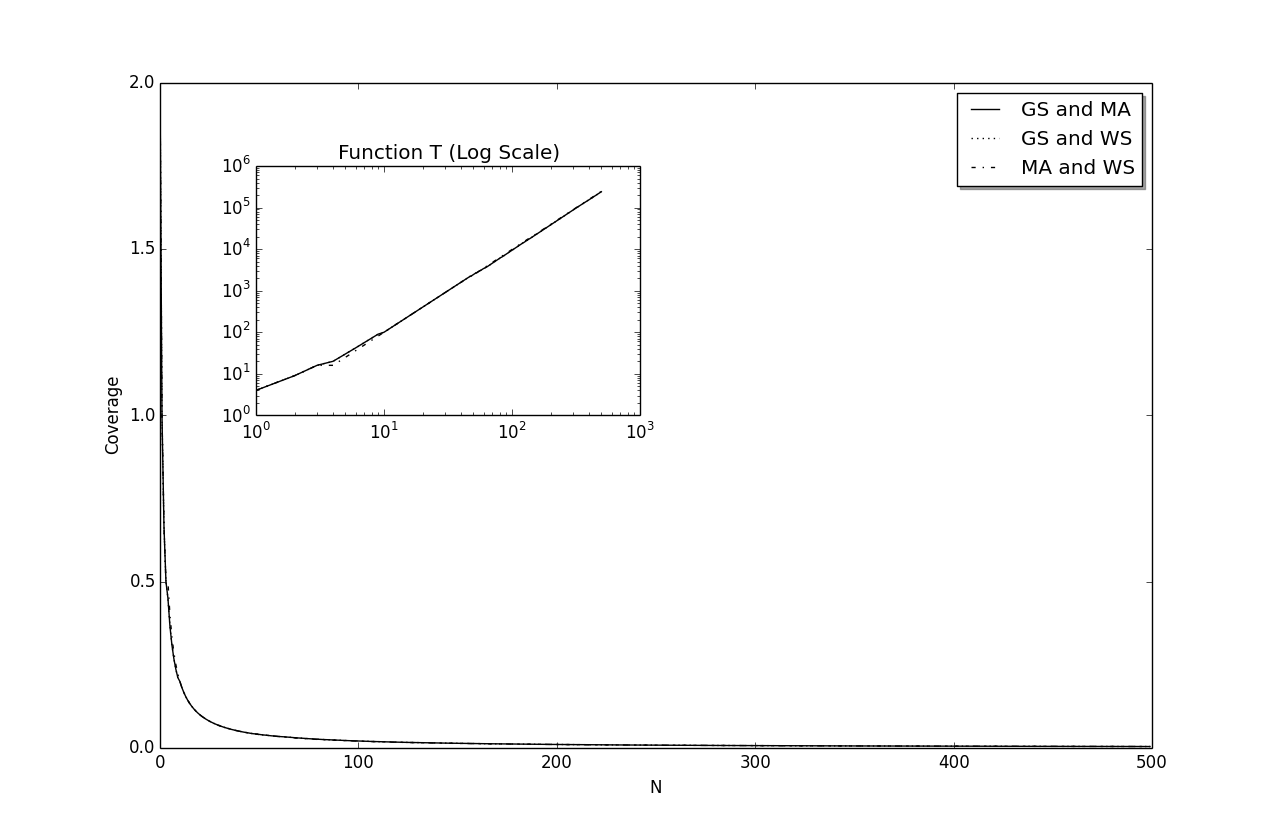}
\caption{\textbf{Keyword: Rechargeable Batteries}. GS, MA and WS are abbreviations for Google Scholar, Microsoft Academic Search and Web of Science respectively. $T$ is quadratic for all pairwise comparisons ($R$ grows so slowly that it is almost constant), hence in the inserted figure it is linear in log scale. As one search further into the results with such a general keyword, one does not get more focused/specialized in the field and thus the coverage approaches zero for increasing $n$.} 
\label{fig:battery}
\end{figure}

The quality of a search depends on how specific the keywords are, for example many disciplines like physics, chemistry and engineering have subfields that research on improving rechargeable batteries. Hence the results from different search engines are drastically different with keywords like ``rechargeable batteries" (Fig. \ref{fig:battery}).

Therefore if a keyword has graphs that is similar to Fig. \ref{fig:battery}, it suggests that one should refine the keyword to be more specific. The keyword is either too ambiguous like ``Phase Transition" and ``Communities Detection", or the topic is studied in many branches of science like ``Genetic Algorithm" and ``Ising Model". In such cases, there is no good stopping rule.

\subsubsection{Type II (1 Local Max and Min)}
\begin{figure}
  \includegraphics[width=1.0\columnwidth]{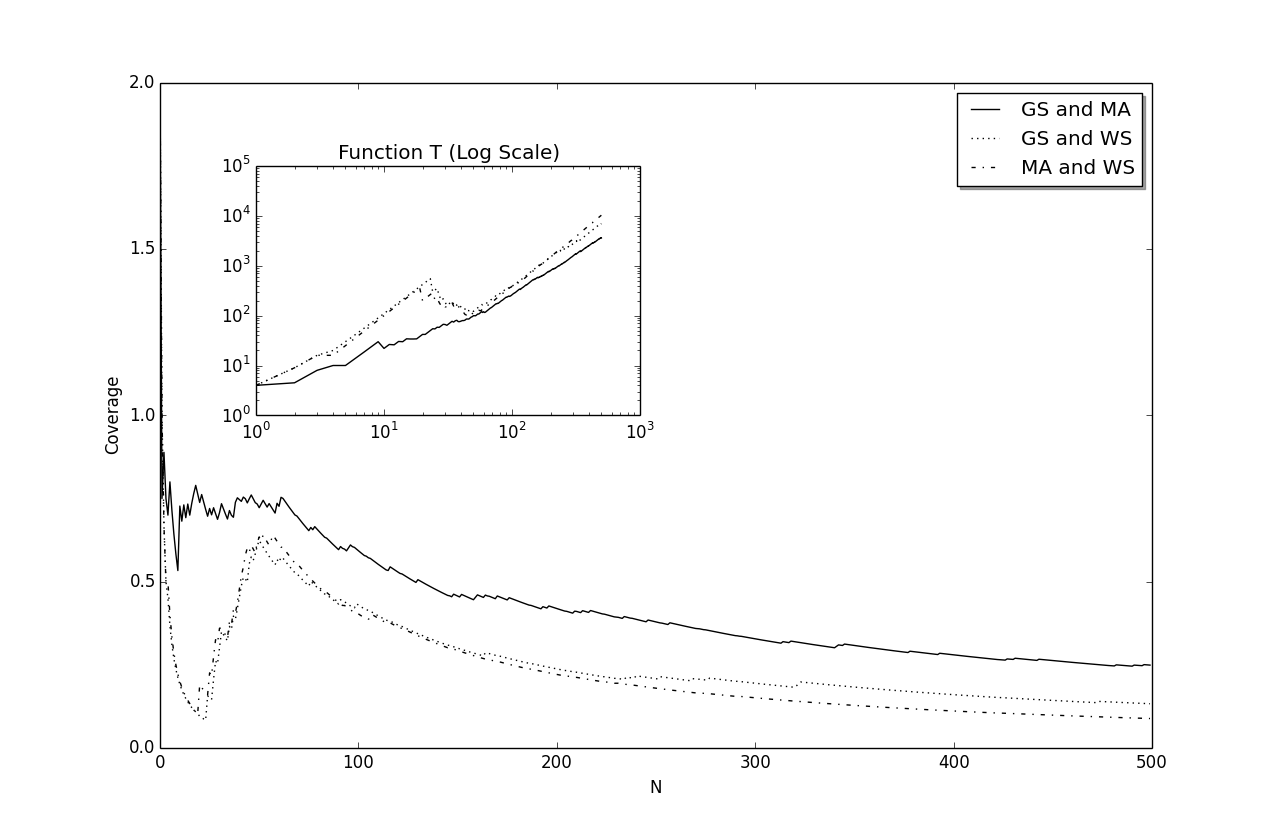}
\caption{\textbf{Keyword: Kauffman Model}. At $n \approx 70$ (local maximum), the rate of change of coverage shifted from zero to negative. This implies one should stop around this point as further search has negative returns. An alternative stopping point is at $n \approx 20$ (local minimum) where it implies that the subsequent articles are already found by the other search engine.} 
\label{fig:kauffman}
\end{figure}

One way to suggest that the search results are drastically different is when $T$ appears to grow quadratically. This usually implies that the choice of keywords is bad and one should discard the search results. However it is not true in general, for example consider the keyword ``Kauffman Model" (Fig. \ref{fig:kauffman}). 

The local minimum of $C$ (coverage) is approximately at $n=20$ where $T$ appears to be linear in log-scale. The rapid increase of coverage plateaued approximately at $n=50$ is the effect that the subsequent articles after $n=20$ in one of the search engines were already listed in the search result of the other search engine. Thus there is little information gain and it is a reasonable to stop at $n=20$.

The local maximum of $C$ plateaued until $n\approx70$, where it is an alternative stopping point for the search. It is an indicator that the search engines' suggestions begin to deviate and hence subsequent articles are less relevant to the keywords. Thus continuing search yields negative returns, which is worse than diminishing returns.

Lastly Fig. \ref{fig:kauffman} also illustrates the sensitivity of the functions $C$ and $T$ with respect to $R$. Since $T$ grows quadratically fast with respect to $n$, the growth of $R$ has to be significant enough to suppress the growth of $T$. Whereas $C$ grows inversely proportionally to $n$, hence minor changes in the dynamics of $R$ can easily affects the growth of $C$. For example the local minimum of $C$ was caused by the change of $R$ at $n\approx 20$, but this change on $T$ is only reflected at $n\approx40$.

Keywords with graphs that are similar to Fig. \ref{fig:kauffman} are unfortunately not very common. Out of the 50 keywords selected for our experiments, only the graphs of ``Kauffman model" and ``Tangled Nature Model"  have both local minimum and maximum.

\subsubsection{Type III (1 Local Min)}
There are many examples that fall into this category, especially for keywords that are less ambiguous and found in very specialized topics. For example ``Skyrmion" has approximated 9000 articles in Google Scholar and most of the publications are also in the database of the other search engines. However every search engines have their own unique algorithms to rank the most relevant articles. 

\begin{figure}
  \includegraphics[width=1.0\columnwidth]{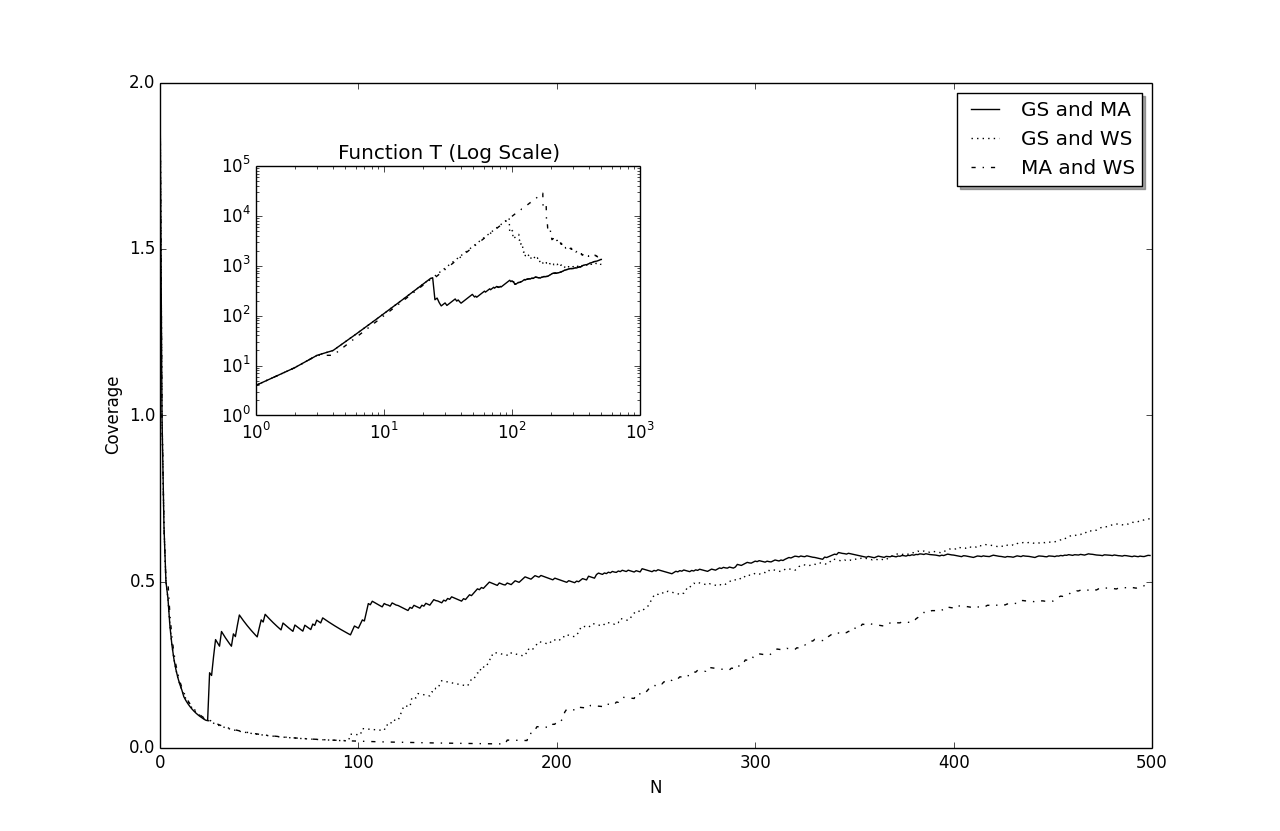}
\caption{\textbf{Keyword: Skyrmion.} In the initially function of $T$, Web of Science and Google Scholar (red) grows quadratically. This implies that their results are significantly different. However at $n \approx 100$, $T$ begins to decrease rapidly and subsequently grows linearly. This implies that the later articles in Web of Science matches the earlier articles in Google Scholar.} 
\label{fig:skyrmion}
\end{figure}

Fig. \ref{fig:skyrmion} shows that the results by the Web of Science initially deviates from Google Scholar and Microsoft Academic Search until $n\approx100$ and $n\approx180$ respectively. After which $T$ converges for all pairwise comparisons. This implies that the initial ordering of ``relevance" by Web of Science is partially the reverse to the result of Google Scholar.

More precisely after the local minimum, the subsequent articles by Web of Science are found in the earlier results of Google Scholar and Microsoft Academic Search. Therefore the coverage increases and there is little information gained. Thus for example if one uses Google Scholar and Web of Science, one should stop the search at $n\approx100$ to avoid diminishing returns as the subsequent articles are mostly found much earlier. This is similar to Type II graphs where one stops at the local minimum, except that the local maximum in Type II graphs makes a stronger case that subsequent articles are divergent.

The following keywords are examples with similar graphs: ``Higgs Boson", ``Spin Fluctuation", ``Resonating Valence Bond", "Rechargeable Lithium-air Batteries", ``Transcatheter Aortic Valve", ``Ocean Acidification", ``Sorting Algorithm", ``Block Cipher", ``Advanced Encryption Standard", ``Evolutionary Dynamics", ``Constantino Tsallis" (search by author) and ``Nonextensive Thermodynamics".

\subsubsection{Type IV (No Significant Feature)}
\begin{figure}
  \includegraphics[width=1.0\columnwidth]{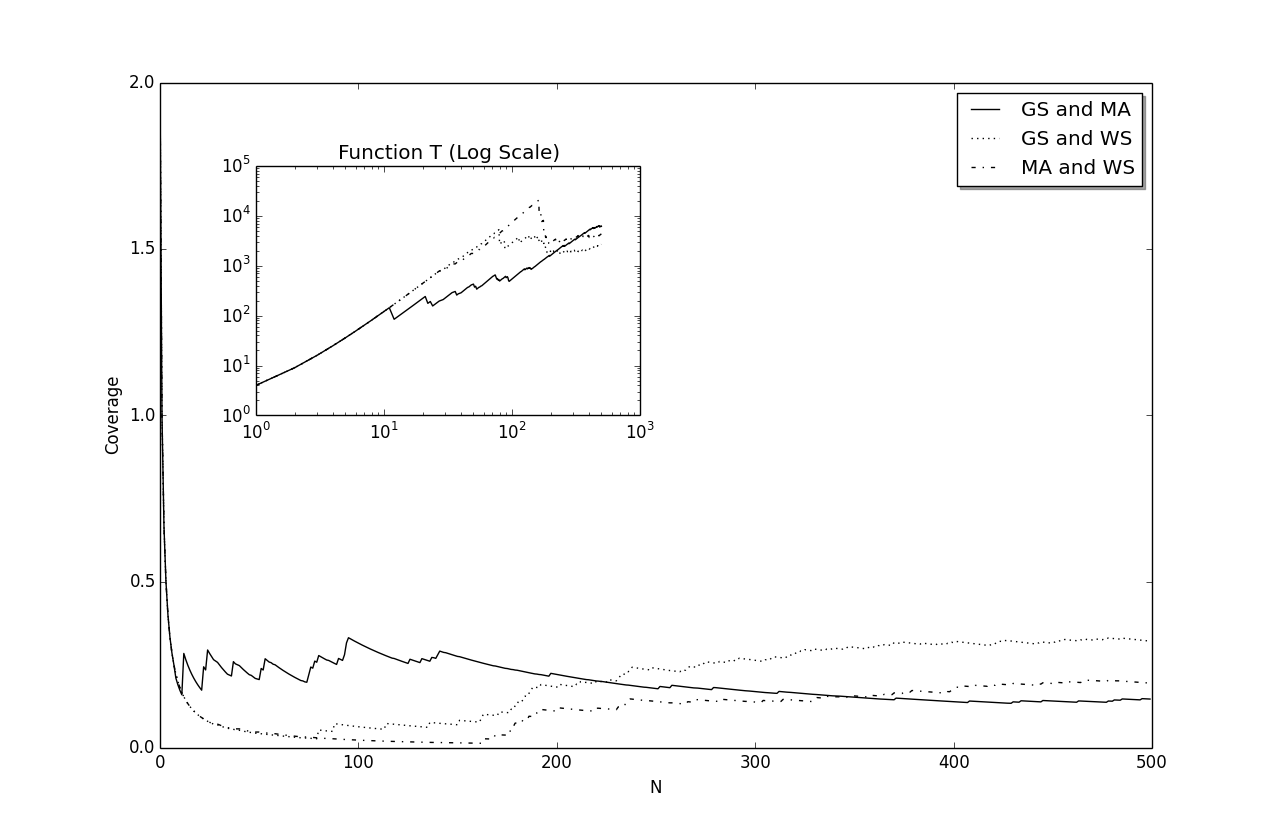}
\caption{\textbf{Keyword: Causality Measures.} There is no significant reference point such that one can suggests a reasonable stop to the search.} 
\label{fig:causality_measures}
\end{figure}

There are many instances where the graphs do not fit into any of the above models due to the nature of the search engines. There is no significant minimum or maximum point for one to suggest a meaningful stop to the search. For example Fig. \ref{fig:causality_measures} is the graph for ``Causality Measures".

We are not able to deduce a general rule to identify keywords that fall into this category: ``Q-Statistics", ``Superconductivity", ``Ant Colony Optimization", `` DNA Methylation", ``Renormalization Group" and ``Hubbard Model". However it appears that the keywords are very specific and the corresponding publications tend to be published in highly specialized journals/conferences. Thus it is possible that there is insufficient data to support a stop for such keywords.

\section{The Centrality Truncated-Ranking of Dynamic Networks}
The order of the results from a search engine is often determined by the relevance of the articles. For instance Google's algorithm has roots from Eigenvector Centrality where it ranks the quality of an article via the behavior of word-of-mouth recommendations. I.e. high ranking articles are either referred by other high ranking articles or by many independent articles.

Therefore the growth of $R$ in essence is also a measure of similarity for the centrality ranking of vertices (search engines ranking). Specifically a linear $R$ with slope 1 indicates high similarity while slow growing (e.g. sublinear) $R$ tends to be less similar. Thus we want to quantify this intuition as a similarity metric between centrality rankings in general. This is closely related to Spearman's Correlation and Kendall-tau Distance as ways to measure the similarity of ranked variables. 

Spearman's Correlation is the variant of Pearson's Correlation for ranked variables where it measures how monotonically two rankings are related. Although it is relevant to our application, the model cannot be used for truncated dataset, i.e comparing the top 100 elements of two rankings. Thus it is not applicable for dynamic systems where the size of the network fluctuates and only the top centrality vertices are interesting. We call such problem the comparison of truncated-ranking.

Kendall-tau Distance measures how likely the order of two elements are agreeable/similar between two rankings. It handles truncated-ranking by ignoring element pairs that does not exists in one of the other rankings. It is sensitive to the ordering of the elements and two rankings are independent (dissimilar) if they are random permutation of each other. This is a good metric until one considers the size of the entire system. It is highly unlikely by random chance in a large system that the top elements of two rankings are similar. Thus even though the ordering of two truncated-rankings might not agreeable in general, its effect is small relative to the fact that the elements between the two truncated-rankings are in common.

\subsection{Measure of Truncated-Ranking Similarities}
The intuition of this metric is based on the observation that when two truncated-rankings are identical, $R$ is a straight line with slope 1 intersecting 0 (i.e. y=x). However when two truncated-rankings are totally dissimilar, i.e. none of the top vertices in one of the ranking is among the top vertices of the other, $R$ is a straight line with slope 0 (i.e. y=0). 

Thus to measure the similarity between two truncated-rankings, this paper proposed the Squared Error difference between $R$ and the line $y=x$. The smaller the Squared Error, the more similar two rankings are. If two rankings do not have the same vertices or the ordering of the vertices are different, then the Squared Error will increase and hence less similar.

Furthermore since the maximum Squared Error is the difference between the lines $y=x$ and $y=0$, we can normalized the measure. Let $N$ be the top $N$ centrality vertices in the network and similarity $S$ is given by:
\begin{equation}\label{metric}
S = 1 - \frac{E(I,R)}{E(I,Z)}, 
\end{equation}
where $I=\{1,2,\ldots,N\}$ is the ideal case (y=x) and $Z=\{0,\ldots,0\}$ ($N$ zeros) is the case where there is no similarity. The Squared Error $E$ is defined as:
\begin{equation}
E = \sum_{j=0}^N |x_j - y_j|^2.
\end{equation}

\subsection{Experiments Methodology}
To simulate a dynamic network that varies in size, we construct a process that adds and removes random vertices from a network in each time step. Between each iteration, the Eigenvector Centrality of the vertices are computed and only the top 1000 vertices are compared. For example let $G_t$ and $G_{t+1}$ be the networks at time $t$ and $t+1$ respectively. If $\bar{v}_t$ and $\bar{v}_{t+1}$ is the ordered list of the top centrality vertices of $G_t$ and $G_{t+1}$ respectively, then $R$ is derived by comparing $\bar{v}_t$ and $\bar{v}_{t+1}$ in the same way as we did with search engines in the section \ref{sec:compareSE}.

We will begin with a network on 10000 vertices constructed using Barab\'{a}si-Albert's construction (See Appendix). In each iteration, $x_r$ random vertices are removed and $x_a$ vertices are added to the network where $x_r$ and $x_a$ are drawn from a normal distribution with mean $1000$ and standard deviation $100$. The new $x_a$ vertices are added into the network using the same mechanism from Barab\'{a}si-Albert's construction.

To further distinguish this metric from Kendall-tau Distance, we will present some special cases in the experiments to demonstrate their differences. Lastly we will measure the similarity of search engines using the real world data in the previous section.

\subsection{Empirical Results}
\subsubsection{Synthetic Network}
Let $Q_1$ and $Q_2$ be two distinct truncated-rankings on the index of vertices of a network, where for instance $Q_1=\{v_a,\ldots,v_z\}$ implies that vertices $v_a$ and $v_z$ are the first and last in the truncated-ranking respectively. From the 1000 iterations in the experiment, the similarity $S$ has a mean of 0.8831 with standard deviation of 0.0697. It is highly correlated to the size of the set $Q_1 \cap Q_2$ with a Pearson's Coefficient of 0.984.

In contrast $S$ is less correlated (Pearson's Coefficient of 0.2443) to Kendall-tau Distance as there are significant changes to the ordering of the top centrality vertices. More importantly the mean Kendall-tau Distance is 0.0332 with standard deviation of 0.0285. This implies that Kendall-tau Distance claims that the two truncated-rankings are dissimilar. The main reason for this dissimilarity is that there are many vertex pairs in one ranking that are not in the ranking of the other. 

For example let $v_i,v_j \in Q_1$ where $v_i$ is ranked higher than $v_j$ in $Q_1$. Suppose $v_i\in Q_2$ and $v_j \not\in Q_2$, then there is neither agreement nor disagreement between $Q_1$ and $Q_2$ on the pair $(v_i,v_j)$. If there are many instances of such pairs, then the Kendall-tau Distance will be close to zero and imply that $Q_1$ and $Q_2$ are independent. However considering the size of the system, it is rare by random chance that there are many common top centrality vertices (e.g. $v_i$). Thus it is counter-intuitive to suggests that the two rankings are not similar.

\subsubsection{Special Cases}
Since $|Q_1 \cap Q_2|$ is highly correlated to our similarity metric $S$, it may appear that $S$ is not insightful. Hence this section presents some special cases of $Q_1$ and $Q_2$ to further distinguish $S$ from the existing metrics.

\textbf{Reverse Ranking:} When $Q_1$ is the reverse of $Q_2$, $|Q_1 \cap Q_2|=1$ and $S=0.7492$. It will be particularly strange to state that the two truncated-rankings are identical given that $|Q_1 \cap Q_2|=1$. Therefore our similarity metric distinguishes itself from the naive approximation of $|Q_1 \cap Q_2|$ by considering the order of the elements in the rankings.

\textbf{Random Permutation:} Suppose $Q_1$ is a random permutation of $Q_2$ and as before it will be strange to assume that both truncated-rankings are identical since $|Q_1 \cap Q_2|=1$. In our simulations on 1000 trials, the mean value of $S$ and Kendall-tau Distance is 0.8993 and -0.0016 respectively. More importantly their Pearson's Correlation Coefficient is 0.9423, thus suggesting that our metric $S$ is similar to Kendall-tau Distance when it comes to the ordering of the rankings. Thus further supports the fact that our metric is more sophisticated than the naive approximation with $|Q_1 \cap Q_2|$.

\textbf{Asymmetry of Ranking:} Unlike the other measures, our metric places more emphasis on the top positions of the truncated-ranking. For example let $Q_1=\{v_a,v_b,\ldots,v_y,v_z\}$, $Q_2=\{v_b,v_a,\ldots,v_y,v_z\}$ and $Q_3=\{v_a,v_b,\ldots,v_z,v_y\}$ where the ``$\ldots$" is identical for all three truncated-rankings. For the other metrics, the similarity between $Q_1$ and $Q_2$ is the same as the similarity of $Q_1$ and $Q_3$. However our metric will state that $Q_1$ and $Q_2$ is less similar than $Q_1$ and $Q_3$.

To empirically support our observation, let $|Q_1 \cap Q_2| = |Q_1|/2 = |Q_2|/2$ where the first halves of $Q_1$ and $Q_2$ are random permutations of each other. Thus there is no common element between the second halves of $Q_1$ and $Q_2$. From 1000 trials, we computed a mean score of 0.8629 and 0.7523 for $S$ and Kendall-tau Distance respectively. Their Pearson's Correlation Coefficient is 0.9573.

If the situation is reversed, i.e. there is no common element between the first halves of $Q_1$ and $Q_2$, and the second halves are random permutation of each other, then the mean score of $S$ and Kendall-tau Distance is 0.3616 and 0.7519 respectively. Since Kendall-tau Distance just counts the number of agreement/disagreement to the element pairs, it does not matter if the missing elements are positioned in the beginning or the end of the ranking. This is different from $S$ as the agreement at the beginning of the rankings has a higher score than the agreement at the end of the rankings.

\subsubsection{Real World Data}
The observation from our real world data (results from search engine) is similar to the results with the synthetic network in the previous experiments. Specifically our metric is positively correlated to the size of $|Q_1 \cap Q_2|$ with a Pearson's Coefficient of $>0.95$ for all pairwise comparisons of the search engines. In addition our metric is almost independent to Kendall-tau Distance with Pearson's Coefficient $\approx -0.1$.

However it is the absolute score of the metrics that is particularly interesting for this section. For instance between the search results of Google Scholar and Microsoft Academic Search, their mean similarity score for $|Q_1 \cap Q_2|$ and Kendall-tau Distance are 0.2799 and 0.0068 respectively. This implies that their results are not similar by those measures. In contrast, our metric has a score of 0.464 with standard deviation of 0.2347. 

Since the score is normalized between 0 and 1, suppose we can let the arbitrary threshold between similarity and dissimilarity to be 0.5. Thus our metric suggests that there is a huge variation between the closeness of the results of Google Scholar and Microsoft Academic Search. This supports the diverging conclusions from other empirical studies that they are \emph{both} similar and dissimilar in general. Therefore our metric is normalized in a way such that it is good for measuring truncated-rankings like search engines' results.

\section{Summary}
Mark-and-Recapture is a simple statistical approximation used by Ecologists to estimate the population size of a species. It can also be used for in applications where one has partial  knowledge of the population. Therefore we proposed using this methodology to assess the completeness of the bibliography of a literature review.

As a proof of concept, we have shown that the approximation is accurate to assess the literature reviews on ``Communities Detection of Graphs/Networks". The estimated number derived using the bibliographies from two literature reviews in 2007 is close to the number of relevant articles (prior to 2008) in the bibliography of a highly cited review paper by Fortunato in 2010.

The concept of measuring the completeness of a bibliography is similar to estimating the proportion of relevant articles found for a given topic. If we assume that the authors of these literature reviews used academic search engines to collect their sources, then it will be useful to assess the completeness of the results returned by the search engines. Thus we reapplied Mark-and-Recapture to study this problem.

The problem has been formulated as a time series (on variable $n$) where the first $n$ articles of different search engines approximates the ratio of the literature found by the search engines to the estimated size of the complete literature. This ratio is known as the coverage of the literature and it is a way to measure the fraction of information known at time $n$. Thus the change of the coverage at time $n$ measures the information gain (or loss) if one is to include the $n^{th}$ article in the research.

Therefore we are able to develop a quantitative stopping criteria for one to follow to maximize his time and resources with the search engine. Lastly the time series also advice us the quality of the choice of keywords used in the search engines. It assumes that the search engines are able to pick the most relevant articles of a given topic and if opinions of these search engines fails to converge, then it indicates that one should refine the choice of keywords.

Finally we show that the same mathematics and ideas can be used to measure the similarity of data-truncated rankings since the problem is parallel to comparing the top articles of search engines. It addresses the issue of truncated ranking in existing similarity metrics like Spearman's Correlation and Kendall-tau Distance. Specifically our metric considers that in a large system, it is unlikely by random chance that there are many common elements found in two different rankings.

In addition in our experiments we showed that the metric is more sophisticated than the cardinality of the intersecting set of two rankings. Not only the metric will penalize the disagreement of the ordering of the rankings, it places more emphasis on the ordering of the top ranks.

A quantitative understanding on the behavior of search and ranking allows us to have a more systematic manner to approach research. Mark-and-Recapture is an approximation to how complete a research is by consolidating the efforts and insights from different sources like literature reviews. However since search engines are now the main source of information, we believed that it will be extremely useful to introduce stopping rules and similarity metrics to study the results from search engines.

\bibliography{database}

\end{document}